# Slicing Virtualized EPC-based 5G Core Network for Content Delivery

Marsa Rayani
Concordia University
Montreal, Canada
m_rayani@encs.concordia.ca

Diala Naboulsi
Concordia University
Montreal, Canada
d_naboul@encs.concordia.ca

Roch Glitho
Concordia University
Montreal, Canada
glitho@encs.concordia.ca

Halima Elbiaze
Université du Québec À Montréal
Montreal, Canada
elbiaze.halima@uqam.ca

*Abstract*— Traditional Content Delivery Networks (CDNs) built with traditional Internet technology are less and less able to cope with today's tremendous growth of content. Information Centric Networks (ICN), a proposed future Internet technology, may aid in remedying the situation. Unlike the current Internet, it decouples information from its sources and provides in-network storage. We expect traditional CDN and ICN-based CDN to co-exist in the foreseeable future, especially as it is now known that it might be possible to evolve traditional CDNs to gain the benefits promised by ICN.  5G providers must therefore aim to offer core network slices on which both ICN-based CDNs and traditional CDNs can be built. These slices could of course also be offered to providers of other applications with requirements similar to those of content delivery. This paper tackles the problem of slicing 5G for content delivery over ICN-based CDNs and traditional CDNs. Only virtualized Evolved Packet Core (EPC)-based 5G is considered. The problem is defined as a resource allocation problem which aims at minimizing the cost of slice assignment, while meeting QoS requirements. An Integer linear programming (ILP) formulation is provided and evaluated in a small-scale scenario.

## I. INTRODUCTION

Online video traffic will show a fourfold increase between 2015 and 2020, according to CISCO [1]. This growth brings new challenges (e.g. scalability, efficiency in content distribution) that traditional Content Delivery Networks (CDNs) [2]) are less and less able to meet. Traditional CDNs consist of geographically distributed replica servers interconnected by IP routers. Their weaknesses are rooted in the fact that they rely on traditional Internet technology [3]. Information Centric Network (ICN) [4] has emerged as a proposed future Internet architecture, and can enable CDNs to meet these challenges. End-users are interested in accessing information, independently of the host where that information is located. ICN decouples information from its source so that information can be located anywhere in the network.

In-network storage (an essential feature of ICN) allows core network routers to cache in addition to the computing capability offered by traditional Internet routers. This makes the deployment of ICN-based CDNs attractive for greenfield CDN providers but very costly for brownfield providers. However, as shown in [5], brownfield operators may also incrementally get most of the gains expected from ICN without radical changes to their traditional Internet infrastructure. We expect CDNs relying on traditional Internet and CDNs relying on ICN to co-exist in the foreseeable future. The emerging 5G can enable this in a flexible and cost-efficient manner through the concept of slicing.

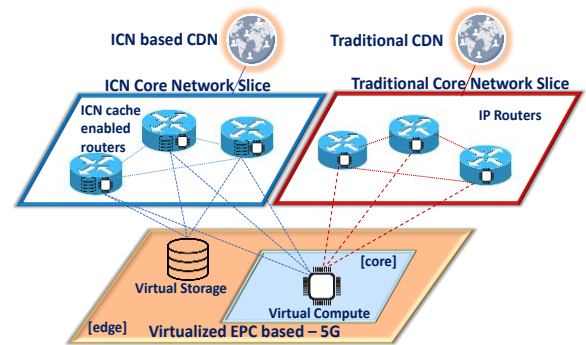

Fig. 1. Virtualized EPC-based 5G

5G network slicing enables the co-existence of different verticals over the same physical infrastructure [6]. This paper proposes a novel scheme in which traditional core networks (i.e. made up of traditional IP routers) coexist with ICN core networks (i.e. made up of cache-enabled routers) on the same 5G physical infrastructure. A virtualized Evolved Packet Core (EPC)-based 5G network [7] is assumed. Furthermore, it is assumed that network function virtualization (NFV) is used as virtualization technology [7].

Using a virtualized EPC-based 5G makes the problem very challenging because EPC routers are not cache enabled. We address this challenge by harnessing the storage offered by EPC servers when we build ICN core network slices. Fig.1 depicts our overall vision: a virtualized EPC-based 5G that offers only virtual computing resources in its core, but virtual storage at the edge; ICN core network slices with cache-enabled routers and traditional IP core network slices with IP routers that rely on the resources provided by the virtualized EPC-based 5G; ICN-based CDN built on the ICN core network slice and traditional CDN built on the IP core network slice. These slices could be used by other applications that have requirements similar to those of content delivery.

Our key contribution is the study of the resource allocation problem for core slice assignment in virtualized EPC-based 5G networks. We aim to minimize slice assignment cost while meeting Quality of Service (QoS) requirements. The rest of the paper is organized as follows. In Sec. II, we discuss the related work. We cover the description and formulation of our problem in Sec. III, and then present the evaluation scenario and results in Sec. IV. Finally, we draw our conclusions and provide future research directions in Sec. V.

## II. RELATED WORK

The problem at hand shares the same space as the general virtual network embedding problem. Researchers have also addressed the specific problem of running virtual networks over a 5G substrate. However, none of the schemes proposed so far is adequate for our problem.

*A. General Virtual Network Embedding Problem*

The general problem consists of embedding virtual networks in a substrate network [8]. However, although we are embedding virtual ICN core networks and virtual IP core networks onto a virtualized EPC-based – 5G, a few peculiarities are worth stressing. First, the embedded ICN core network nodes have computing and storage capabilities. They thus need to be mapped onto two distinct nodes in the infrastructure: a compute node in the core of the substrate network and a storage node at the edge. Second, the in-network caching capability imposes requirements on the QoS for users accessing the content stored there. These requirements need to be accounted for in the mapping. Third, the substrate network uses NFV as its virtualization technology. Unfortunately, the current literature does not address these specifics in a satisfactory manner. Reference [9] for instance considers embedding nodes with computing and storage capabilities. However, it assumes that the substrate nodes have both computing and storage capabilities. Authors of [10] do tackle the 1-to-many mapping issue, but the mapping is done for parallelization purposes.

*B. Virtual Networks Over a 5G Substrate*

As mentioned in Sec. II.A, embedding ICN virtual networks is rather challenging and several researchers have tackled the issue. In [11] for instance, authors propose a general 5G-ICN architecture, a detailed network slicing architecture for the embedding, and deployment models. However, the paper does not cover the resource allocation problem. Furthermore, it assumes a 5G substrate core network that provides computing and storage, while we consider storage is offered at the edge and computing is offered in the core of a virtualized EPC-based 5G. Reference [12] proposes an architecture in which a virtual wireless ICN network and a virtual traditional wireless network co-exist on the same physical 5G wireless network. A scheme is proposed to jointly optimize caching and resource allocation. However, it handles resource allocation separately for individual users. This same work also assumes a 5G substrate core network which provides both computing and storage. In addition, their substrate network is made up of real resources, while in our case it is made up of an NFV Infrastructure (NFVI).

While in the previous paragraph, the researchers all focus on ICN, there are also some studies that address the problem at large. The focus of these works so far is on how to jointly optimize resource and revenue, as in [13]. There, an auction model is proposed to solve the problem. However, the work does not consider the peculiarities of embedding ICN slices.

III. RESOURCE ALLOCATION PROBLEM

We now present the problem of resource allocation for core slices assignment in virtualized EPC-based 5G networks. We state the problem in Sec. III-A, present our system model in Sec. III-B, and the problem formulation in Sec. III-C.

*A. Problem Description*

We aim at allocating resources for a set of traditional and ICN core network slices, over a substrate network formed by a virtualized EPC-based 5G network. We assume that NFV is used as a virtualization technology. Therefore, the virtualized EPC-based 5G substrate network represents an NFVI. It offers virtualized computing resources in the core of virtualized EPC-based 5G and virtualized storage resources through the edge of virtualized EPC-based 5G.

Each core network slice is formed by a set of Virtual Network Functions (VNFs). A traditional core network slice includes only traditional IP routers, and thus includes only compute VNFs. An ICN core network slice instead includes cache-enabled routers, capable of caching content. Each cache-enabled router is formed by one computing and one storage VNF. By that, an ICN slice includes a mixture of computing and storage VNFs. Our objective is to enable the embedding of slices to the 5G substrate network at the lowest operational cost, while still meeting the QoS requirements.

*B. System Model*

Here, we present our system model.

*1) 5G Substrate Network*

We represent the 5G substrate network as an undirected graph $IG = (N, L)$ where $N$ is a set of nodes, with each node $n$ representing a virtualized computing resource at the core or storage resource at the edge in the substrate network and $L$ is a set of edges linking them. An edge $(n, n') \in L$ linking nodes $n$ and $n'$ represents a logical communication link between them. Each node $n$ has a type $t_n \in T$, where $T = \{compute, storage\}$. We use $R_n$ to refer to the resource capacity of node $n$. We employ $c_n$ and $c_{nn'}$ to denote the cost of one resource unit at node $n$ and one unit of network bandwidth over the edge $(n, n')$. We use $B_{nn'}$ and $d_{nn'}$ to represent the bandwidth capacity and delay of edge $(n, n')$.

*2) Slices*

We define $S$ as the set of traditional and ICN core network slices to map to the substrate network. Each slice $s \in S$ includes a set of VNFs $V_s$ to map to the substrate network nodes. We associate to a slice $s \in S$ a set of users' demands $U_s$ and a set of replica servers $W_s$ that are given. We build a graph $SG_s = (H_s, E_s)$ to represent slice $s$. There, $H_s$ is a set of nodes defined as $H_s = V_s \cup U_s \cup W_s$ and $E_s = E_s^V \cup E_s^U \cup E_s^W$ is a set of edges that link nodes in $H_s$. $E_s^V$ are edges that exist between the VNFs in slice $s$. $E_s^U$ are edges that exist between users and VNFs in slice $s$. $E_s^W$ are edges that exist between VNFs and CDN surrogate servers. A VNF $v \in V_s$ has a type $t_v \in T$. A traditional core network slice has VNFs of type computing only. Instead, as an ICN slice relies on cache-enabled routers, it has VNFs of computing and storage types. We use $r_v$ to refer to the resource demand of a VNF $v \in V_s$. Each link $(v, v') \in E_s$ requires $b_{vv'}$ bandwidth units.

*C. Problem Formulation*

We formulate our problem as an ILP problem. We define the following decision variables.

$$x_{nv} = \begin{cases} 1, & \text{if } v \in V_s \text{ is located at } n \in N \\ 0, & \text{otherwise} \end{cases}$$

$$y_{vv'}^{nn'} = \begin{cases} 1, & \text{if } (v, v') \in E_s \text{ is mapped to } (n, n') \in L \\ 0, & \text{otherwise} \end{cases}$$

**Operational cost:** The operational cost has two components.

*Deployment cost* ($C^{dep}$): It represents the cost of resources over infrastructure nodes, allocated for VNFs:

$$C^{dep} = \sum_{s \in S} \sum_{n \in N} \sum_{v \in V_s \mid t_v = t_n} c_n r_v x_{nv}$$

*Communication cost* ($C^{com}$): It represents the cost of network bandwidth consumed in the communication in the core slices:

$$C^{com} = \sum_{s \in S} \sum_{(v,v') \in E_s} \sum_{(n,n') \in L} c_{nn'} b_{vv'} y_{vv'}^{nn'}$$

Our objective is to minimize the overall operational cost:

$$Min\ (C^{dep} + C^{com})$$

**Constraints:** The following constraints are considered in our problem. Each VNF instance should be mapped to one infrastructure node, as indicated in constraint (1):

$$\sum_{n \in N \mid t_v = t_n} x_{nv} = 1\ ;\ \forall v \in V_s, s \in S \quad (1)$$

Constraint (2) ensures that the number of VNFs placed over an infrastructure node does not exceed its capacity:

$$\sum_{s \in S} \sum_{v \in V_s \mid t_v = t_s} r_v . x_{nv} \leq R_n\ ;\ \forall n \in N \quad (2)$$

Constraint (3) imposes that the bandwidth capacity of an infrastructure link is not exceeded:

$$\sum_{s \in S} \sum_{(v,v') \in E_s} b_{vv'} y_{vv'}^{nn'} \leq B_{nn'}\ ;\ \forall (n,n') \in L \quad (3)$$

We use a matrix $P$ to identify the location of users' demands, and a matrix $Q$ to identify the location of surrogate servers:

$$p_{nv} \in P = \begin{cases} 1, & \text{if } v \in U_s \text{ is located at } n \in N \\ 0, & \text{otherwise} \end{cases}$$

$$q_{nv} \in Q = \begin{cases} 1, & \text{if } v \in W_s \text{ is located at } n \in N \\ 0, & \text{otherwise} \end{cases}$$

We thus link variables $x_{nv}$ and $y_{vv'}^{nn'}$ for edges in $E_s^U$ and $E_s^W$:

$$y_{vv'}^{nn'} = x_{nv} p_{n'v'}\ ;\ \forall s \in S, (v,v') \in E_s^U, (n,n') \in L \quad (4)$$

$$y_{vv'}^{nn'} = x_{nv} q_{n'v'}\ ;\ \forall s \in S, (v,v') \in E_s^W, (n,n') \in L \quad (5)$$

We link the variables $x_{nv}$ and $y_{vv'}^{nn'}$ for edges $(v,v')$ in $E_s^V$ with the following constraint:

$$y_{vv'}^{nn'} = x_{nv} x_{n'v'}\ ;\ \forall s \in S, (v,v') \in E_s^V, (n,n') \in L \quad (6)$$

Constraint (6) is not linear, we linearize it as follows:

$$y_{vv'}^{nn'} \leq x_{nv}\ ;\ \forall s \in S, (v,v') \in E_s^V, (n,n') \in L \quad (7)$$

$$y_{vv'}^{nn'} \leq x_{n'v'}\ ;\ \forall s \in S, (v,v') \in E_s^V, (n,n') \in L \quad (8)$$

$$y_{vv'}^{nn'} \geq x_{nv} + x_{n'v'} - 1\ ;\ \forall s \in S, (v,v') \in E_s^V, (n,n') \in L \quad (9)$$

We consider a maximum latency $D_{max}$ imposed to meet users' demands. For an ICN core slice, the threshold needs to be met when serving a user from a storage VNF or from the surrogate server. For a traditional core slice, the threshold needs to be met when serving users from the surrogate server. To reflect this, we define $SG_{s'} = (H_{s'}, E_{s'})$ as a subgraph of $SG$ with $H_{s'} \subset H_s$ and $E_{s'} \subset E_s$, such that $(v,v') \in E_{s'}$ only if $v \in H_{s'}$, and $v' \in H_{s'}$. $SG_{s'}$ represents the flow from one cache or from the original server to a set of users. We define a set of sub-slices that map to these subgraphs as $S'$, with $s' \in S'$ representing one subgraph. The maximum latency is then imposed with constraint (10):

$$\sum_{(v,v') \in E_{s'}} \sum_{(n,n') \in L} d_{nn'} . b_{vv'} . y_{vv'}^{nn'} \leq D_{max}\ ;\ \forall s' \in S' \quad (10)$$

IV. PERFORMANCE EVALUATION

This section presents our evaluation scenario and results.

*A. Evaluation Scenario*

We describe below our evaluation scenario.

**5G Substrate Network:** We build on the 4G EPC network in [14] to construct our 5G substrate network. We consider the 4G EPC functional entities serving the west side of the USA. These include one Packet Data Network Gateway (P-GW) and three Serving Gateways (S-GWs), offering storage resources at the edge. We assume a set of core routers, offering computing resources, exists among the functional entities, as indicated in Fig.2. We further assume two sets of users, located in different cities, and surrogate servers serving them, as highlighted in the figure. Each infrastructure node is located in a different city, and so we use the hourly electricity price for the corresponding location as the cost of infrastructure resources [15]. These prices range from 0.0833 to 0.1776 $/kWh. We assume all functional entities hold the same storage capabilities and make the same assumption for all routers regarding their computing capabilities. For links among infrastructure nodes, we consider the latency to map to the average delay of ping time obtained from global statistics data [16]. We assume that each edge has 10 Gbps of bandwidth capacity and further consider that the bandwidth unit cost for all edges is equal to 0.155$/GB.

**Slices:** We consider the presence of two ICN and 5G core slices. The slices allow each to serve a set of users and account for the presence of surrogate servers. For the ICN slice, the VNFs types are both storage and computing. For the 5G slice, only computing VNFs are assumed as indicated in Fig.3

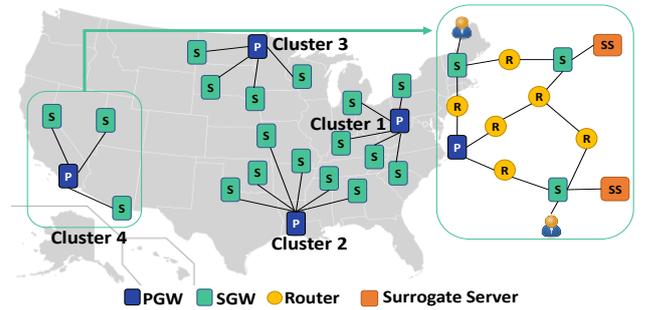

Fig. 2. 5G substrate network

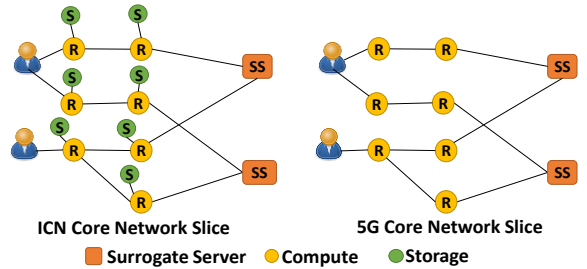

Fig. 3. ICN core network slice and 5G core slice network

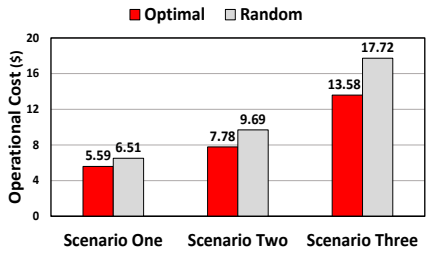
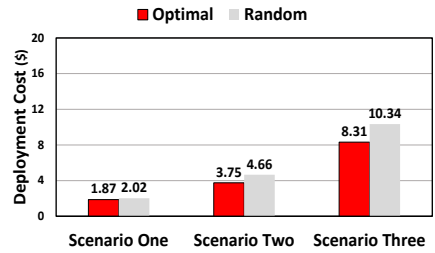
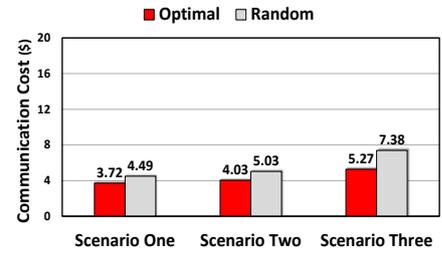

Fig. 4(a). Overall Cost  Fig. 4(b). Deployment Cost  Fig. 4(c). Communication Cost

Fig. 4. Evaluation results

TABLE I. SIMULATION SCENARIOS

|  | Scenario one | Scenario two | Scenario three |
| --- | --- | --- | --- |
| VNF compute | 7 ICN +7 5G | 14 ICN +14 5G | 28 ICN +28 5G |
| VNF storage | 7 ICN | 14 ICN | 28 ICN |

Multiple scenarios with different numbers of VNFs are considered, as detailed in Table I. We further assume that a VNF requires 10% of the capacity of an infrastructure node. We consider that 1 Gbps of bandwidth is required for communication between two VNFs. We set the QoS threshold to 1555 ms.

### B. Evaluation Results

We solved the problem over the scenarios described in the previous subsection, using IBM CPLEX solver. In Fig.4(a), we show the resulting overall cost. As can be observed, the overall cost grows as the number of VNFs in the slices increases. More importantly, we compare the overall cost to that obtained based on a random placement of VNFs that still satisfies our maximum latency threshold and infrastructure node capacities. We can see that there are notable differences between the two, with a random placement leading to higher costs than the optimal solution derived by our model. This is especially the case for the third scenario with the largest number of VNFs.

Fig.4(b) and 4(c) further show the deployment and communication costs for both the optimal solution and the randomly generated solution. Similarly, we notice notable differences between the random and optimal mappings. Overall, these observations highlight the importance of considering a strategic mapping of slices over the infrastructure, for an efficient management of resources.

## V. CONCLUSION

In this paper, we studied the problem of mapping core slices for content delivery to a virtualized EPC-based 5G network. We considered core network slices where ICN-based CDN and traditional CDN could be built. Our objective was to allocate resources at minimum cost, while meeting QoS requirements over each slice. We formulated the problem as an ILP problem. We evaluated it and compared its outcome to a random solution. Our results show that a random solution can result in higher costs than the optimal solution, which indicates the need for adequate resource allocation algorithms. In the future, we plan to extend this work by designing efficient resource allocation algorithms. We further aim to go beyond the core network, and consider the 5G radio access network as well.